\documentclass[aps,prd,twocolumn,epsf,superscriptaddress]{revtex4}

\usepackage{amsfonts}
\usepackage{amsmath}
\usepackage{amssymb,amsbsy}
\usepackage{bm}
\usepackage{makeidx}
\usepackage{latexsym}
\usepackage{graphicx}
\usepackage{epsfig}
\usepackage{subfigure}
\usepackage{dsfont}
\usepackage{mathcomp}
\usepackage{color}
\usepackage{mathrsfs}
\usepackage{epsf}
\usepackage{caption}
\usepackage{pstricks}
\newcommand{\beq}{\begin{eqnarray}}
\newcommand{\eeq}{\end{eqnarray}}
\newcommand{\be}{\begin{equation}}
\newcommand{\ee}{\end{equation}}

\newcommand{\gapp}{\mathrel{\raise.3ex\hbox{$>$}\mkern-14mu
              \lower0.6ex\hbox{$\sim$}}}
\newcommand{\lapp}{\mathrel{\raise.3ex\hbox{$<$}\mkern-14mu
              \lower0.6ex\hbox{$\sim$}}}

\newenvironment{Figure}
  {\par\medskip\noindent\minipage{\linewidth}}
  {\endminipage\par\medskip}



\begin{document}
\title{Electroweak Vacuum Stability in light of BICEP-2}
\author{Malcolm Fairbairn and Robert Hogan}

\affiliation{Physics, Kings College London, Strand, London WC2R 2LS, UK}

\begin{abstract}
\noindent
We consider the effect of a period of inflation with a high energy density upon the stability of the Higgs potential in the early universe. The recent measurement of a large tensor-to-scalar ratio, $r_T \sim 0.16$, by the BICEP-2 experiment possibly implies that the energy density during inflation was very high, comparable with the GUT scale. Given that the standard model Higgs potential is known to develop an instability at $\Lambda \sim 10^{10}$ GeV this means that the resulting large quantum fluctuations of the Higgs field could destabilize the vacuum during inflation, even if the Higgs field starts at zero expectation value.  We estimate the probability of such a catastrophic destabilisation given such an inflationary scenario and calculate that for a Higgs mass of $m_h=125.5$ GeV that the top mass must be less than $m_t\sim 172$ GeV.  We present two possible cures: a direct coupling between the Higgs and the inflaton and a non-zero temperature from dissipation during inflation. 
\end{abstract}

\maketitle

The discovery of the (Brout-Englert-)Higgs boson of the standard model has rightly been heralded as one of the most significant scientific discoveries of recent years \cite{ATLAS1207,CMS1207}.  At present there is no evidence to suggest that the particle is anything other than a fundamental scalar field \cite{Ellis1303} and there is not yet any evidence for the existence of other particles beyond the standard model of particle physics \cite{ATLAS-CONF-2014-006}.

The observed values of the standard model parameters, in particular the top mass and the Higgs mass, imply that the running of the Higgs quartic coupling $\lambda_h$ may be such as to become negative at large values of the Higgs field $h$ \cite{degrassi1205,buttazzo1307}.  For example, for the central value of $m_{t}=173.34$ GeV and $m_{h}=125.66$ GeV the Higgs potential becomes unstable at a scale $\Lambda$ just above $h=10^{10}$ GeV.  For this reason there has been much attention paid to the tunneling rate from our vacuum to the unstable vacuum in order to put bounds on the lifetime of our metastable minimum at $h=246$ GeV \cite{Isidori0104,Isidori0712,ArkaniHamed0801,Ellis0906,EliasMiro1112,degrassi1205,buttazzo1307}.

In this letter we will only consider models containing General Relativity and field theory with minimal couplings between the two, in which case the Higgs field acting alone does not seem to be a good inflationary candidate.  Nevertheless, since we now know it exists, the behaviour of the Higgs field during inflation has been frequently considered before \cite{Espinosa0710} and since its discovery \cite{DeSimone1208,Lebedev1210,Enqvist1211,Enqvist1306,Enqvist1310,Kobakhidze1301}. During inflation all fields lighter than the Hubble rate $H$ will receive stochastic quantum fluctuations of order $H/{2\pi}$ per Hubble time from the Gibbons-Hawking temperature but scalar fields (and gravitons) in particular can undergo anomalous growth when the wavelength of some Fourier mode exceeds the de Sitter horizon \cite{Finelli0808}.  Since successful inflation with generation of scalar perturbations which fit the data well can be achieved for a wide variety of inflationary energy scales, the magnitude of these quantum fluctuations can be relatively small.

Very recently results were presented to the community from the BICEP-2 experiment concerning measurements of the polarisation of the cosmic microwave background radiation \cite{Ade1403}.  While the results require verification, the observations seem to be most consistent with a tensor-to-scalar ratio of around $r_T=0.16^{+0.06}_{-0.05}$ for what they claim is their most realistic dust model.  If one chooses to interpret this result as being due to gravitational waves produced during inflation, it immediately sets the scale of the energy density during inflation to be very large, around the GUT scale, $10^{16}$ GeV.  Under this assumption, there are a number of inflationary models that previously seemed to be under pressure from WMAP and Planck constraints on $r_{T}$ which suddenly become viable once more \cite{WMAP9,Planck}.  The simplest such model is the quadratic inflationary potential $V=\frac{1}{2}m^2\phi^2$.

No matter what the shape of the potential, if the energy density during inflation is as high as the GUT scale, then the Higgs field will receive stochastic fluctuations which will push it to expectation values typically of order $10^{13}$ GeV or higher during the 50-60 efolds of inflation, even if one assumes that its value at the beginning of those final efolds was zero (for a previous detailed analysis of this problem see \cite{Espinosa0710}).  If the Higgs field has instabilities above $10^{10}$ GeV the tunnelling calculation would therefore be rendered irrelevant - at or before the end of inflation the Higgs field will roll classically into the unstable minimum leading to a Universe incompatible with the one we live in (in which at the {\it very} least the particle physics would be very different from what we observe).

We therefore know that in such a scenario, some physics must be responsible for the Higgs field not rolling into the unstable minimum.  This could be achieved in ways that have nothing to do with the inflaton field, for example through couplings to particles which have not yet been discovered which enter the running and prevent the quartic coupling from destabilising \cite{EliasMiro1203}, through a non-minimal coupling of the Higgs to gravity \cite{Espinosa0710}, or through the actual top mass and Higgs mass being such that our minimum at $h=246$ GeV is the true minimum of the theory (although such a scenario is currently incompatible with experiment at $>2\sigma$).  

 Another point of view is that the solution to this problem could lie with the inflaton itself and may be the first clue concerning its couplings to standard model particles.  One way in which the inflaton could affect the problem in a non-dynamical way would be to consider the coupling of the inflaton to the Higgs field and calculate how this affects the running of the quartic coupling.  For values of the coupling between the Higgs and the inflaton of order $10^{-1}$ that the entire Higgs potential becomes stable up to the Planck scale, however the running of the quartic coupling due to the inflaton would only change above the mass scale of the inflaton, and for many simple models the energy scale favoured by BICEP-2 would point to an inflaton mass close to $10^{13}$ GeV, above the energy scale required to stabilise a quartic coupling which becomes negative at $10^{10}$ GeV.

There are two other ways that the inflaton could affect the vacuum stability - the first is the situation where the Higgs has a sufficiently large direct coupling to the inflaton itself such that the instability is prevented from appearing during inflation when the inflaton has as large vacuum expectation value \cite{Lebedev1210,Kobakhidze1301}.  The second way is through dissipative effects during inflation which could create temperature corrections to the Higgs mass which would result in it rolling into the standard model minimum at $h=246$ GeV.

In this paper we will consider the best case scenario where the Higgs is sitting at the origin 50 efolds before the end of inflation. Of course, if the total number of efolds is much greater than those required to solve the horizon problem then the probability of collapse will only increase. The argument that the Higgs field may roll in from the Planck scale just as the inflaton field does carrys weight but here we choose to be conservative.  In any case if the initial conditions are such that the $h> \Lambda$ at the beginning of inflation then, without additional contributions to the Higgs potential, the Higgs will just classically roll into the instability, perhaps before the inflaton potential comes to dominate the energy density of the Universe. 

The stochastic growth of fluctuations of the Higgs field during inflation can be described by \cite{Finelli0808}
\begin{equation}
\frac{d \langle h^2 \rangle}{d t}+\frac{2}{3 H(t)}\langle h V_{\text{eff}}^\prime \rangle  =\frac{H^3(t)}{4 \pi^2},
\label{eqn: stochastic}
\end{equation}
where $H(t)$ is the Hubble constant during inflation and $V_{\text{eff}}$ is the effective potential. We see that a large positive effective mass will damp these fluctuations while $H(t)$ acts as to source them. For the Higgs field we have
\begin{equation}
V_{\text{eff}}(h)=\frac{1}{4	}\lambda_{\text{eff}}(h) h^4,
\end{equation}
where the $\lambda_{\text{eff}}$ contains the 2-loop effective potential corrections and runs with energy scale according to 3-loop RGEs \cite{degrassi1205,buttazzo1307}. As discussed earlier, the running is such that $\lambda_{\text{eff}}$ can become negative at $\Lambda \sim 10^{10} $ GeV  due to large negative contributions of the top Yukawa to the beta function. Using \cite{degrassi1205,buttazzo1307} we define our inputs at $m_{t}$ and extrapolate up to large scales. The resulting $\lambda_{\text{eff}}$ for the central values is shown in figure \ref{leff}.

\begin{Figure}
\centering
\includegraphics[width=1.0\linewidth]{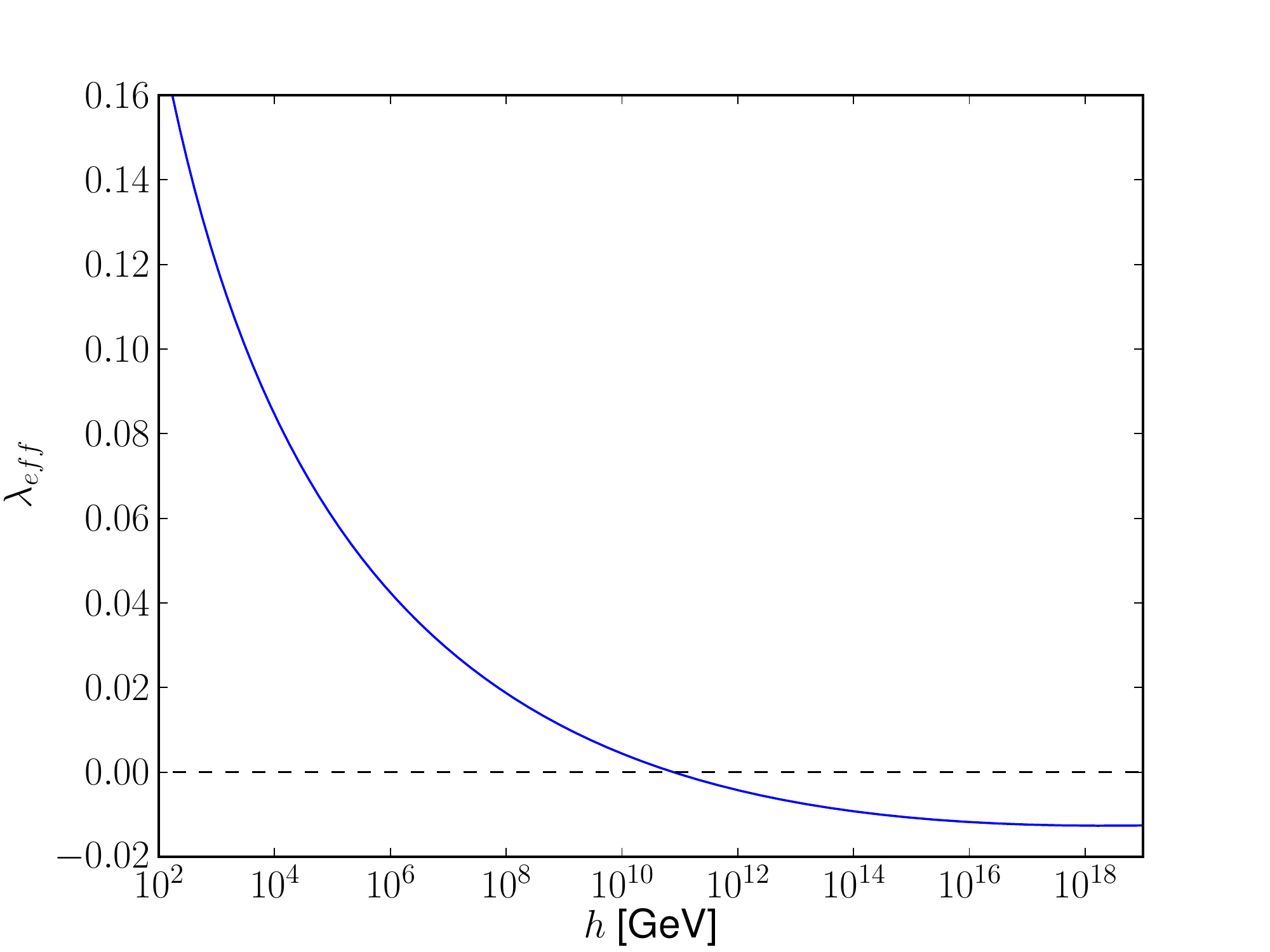}
\captionof{figure}{\it The running of the renormalized effective coupling for $m_h=125.66$ GeV and $m_{top}=173.34$ GeV. An instability develops when $\lambda_{\text{eff}}$ crosses zero.}\label{leff}
\end{Figure}

The solution to  (\ref{eqn: stochastic}) at $t=t_{\text{end}}$ gives the variance of probability distribution of $h$ at the end of inflation. The probability of fluctuating into the unstable region is then given by the fraction of this distribution with $h> \Lambda$.  This calculation of the probability is slightly complicated by the fact that there is a different value of the Higgs field $h$ in each Hubble volume during inflation.  The approach that we use therefore is to evaluate at each efold of inflation the probability in a single Hubble volume and weight that with the number of independent Hubble volumes at each efold which end up within our horizon.  This can be written as
\begin{equation}
P_{surv}=\prod_{N=1}^{N_{efolds}}\left[1-\int_\Lambda^\infty\sqrt{\frac{2}{\pi\langle h^2\rangle_N}}\exp\left(-\frac{1}{2}\frac{h^2}{\langle h^2\rangle_N}\right)\right]^{j_N}
\end{equation}
where $N$ is the number of efolds, $\langle h^2\rangle_N$ is the variance of $h$ after $N$ efolds evaluated using equation (\ref{eqn: stochastic}) and $j_N$ is the number of separate Hubble volumes at efold $N$ which end up within our past light cone today.  For a more rigorous calculation of this probability, see \cite{Espinosa0710}.  At this stage the reader might worry that it is possible to fluctuate to values of $h$ greater than $\Lambda$ before fluctuating backwards without the negative energy density Higgs potential significantly changing the overall positive energy density of inflation.  We have looked numerically at the behaviour of the Higgs field introducing stochastic perturbations in an ad hoc way and when $h$ does find itself above $\Lambda$ it does rapidly evolve towards classically running to large positive values of $h$ deep within the de-Sitter vacuum.  Still it would be nice to perform a more detailed calculation of these rather complicated dynamics.

To demonstrate the effect of these fluctuations we adopt the simplest potential which could be compatible with the BICEP-2 results, $V$=$\frac{1}{2}m_\phi^2\phi^2$ which fits $r_T$ for the Planck favoured value for the spectral index of $n_s=$0.96 with a value of $\phi$ corresponding to the largest observable scales in the CMB of $\sqrt{200}M_{Pl}$, where $M_{pl}$ is $\sqrt{1/8\pi G}$, resulting in 50 efolds of inflation (see e.g. \cite{Ellis1312}).  The amplitude of the scalar perturbations requires the mass of the inflaton field to be $m_\phi=5.97\times 10^{12}$ GeV.

In figure \ref{without} we plot the probability of the Higgs field not finding itself above the instability scale after 50 efolds of inflation.
\begin{Figure}
\centering
\includegraphics[width=1.1\linewidth]{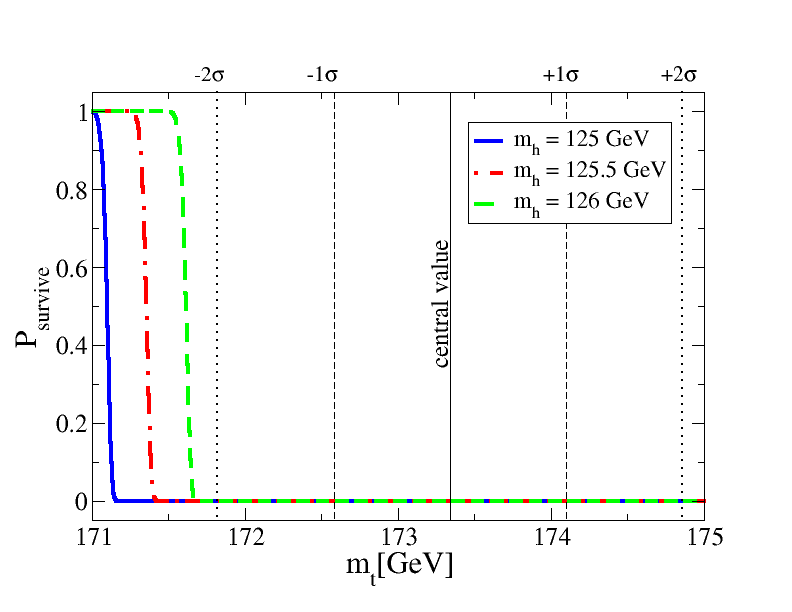}
\captionof{figure}{\it The probability of the Higgs field not ending up above the instability scale $\Lambda$ in any of the Hubble volumes in our past horizon during inflation as a function of top mass $m_t$.  We plot the results for three values of the Higgs mass $m_h$.  We also plot the 1$\sigma$ and 2$\sigma$ limits on $m_t$.}\label{without}
\end{Figure}
Because of the sensitivity of the value of $\Lambda$ to the top mass, we believe that the final results presented in this paper for the values of $\lambda_{\phi h}$ and $T$ required to prevent instability would be robust even if the probability was calculated in a different way (to within about an order of magnitude).  Different ways to evaluate the probability may lead to different constraints on the allowable top mass values, but we believe that the central value of $m_t$ being incompatible with stability during this kind of inflation is a robust statement.

Figure \ref{without} shows that naively, without further physics, inflationary fluctuations would push the Higgs field over the top of the potential to above the critical scale $\Lambda$ for the best favoured values of the Higgs mass and top-quark mass into a (presumably anti de-Sitter) vacuum. This is clearly incompatible with our Universe so it must be rectified.  This amounts to altering $V_{\text{eff}}$ in some way such that the probability of collapse is decreased to an acceptable level. We would like to demonstrate two phenomenological possibilities and show how they could both in principal work. 

Firstly, perhaps the simplest solution (considered already in \cite{Lebedev1210,Kobakhidze1301}) is to introduce a direct coupling between the Higgs and the inflaton such that,
\begin{equation}
V_{\text{eff}} \rightarrow V_{\text{eff}}  +\frac{1}{2} \lambda_{\phi h} \phi^2 h^2.
\end{equation}
During inflation, $\phi$ has a large value so this contribution could stabilise the vacuum completely for suitable $\lambda_{\phi h}$, or at least push $\Lambda$ to larger scales and reduce the probability of collapse if the Higgs starts with a low expectation value. 

\begin{Figure}
\centering
\includegraphics[width=1.1\linewidth]{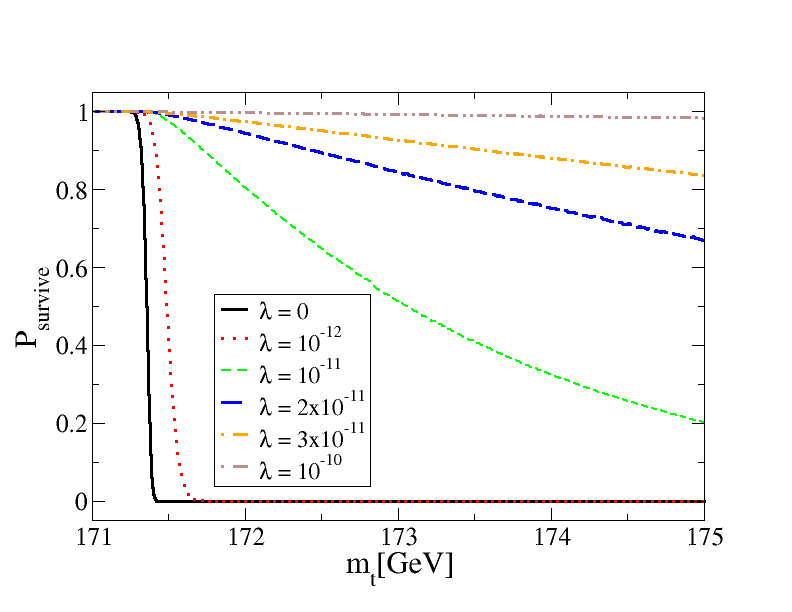}
\captionof{figure}{\it The probability of the Higgs field not ending up above the instability scale $\Lambda$ in any of the Hubble volumes in our past horizon during inflation as a function of top mass $m_t$.  We plot the results for different values of the Higgs-inflaton coupling mass $\lambda_{\phi h}$.}\label{coupling}
\end{Figure}

The effect of such a direct coupling is shown in figure \ref{coupling} where we find that for $\lambda_{\phi h}$ a few $\times 10^{-11}$ the modified effective mass of the Higgs is such that probability of surviving until the end of inflation increases dramatically.

In the absence of a sufficiently large direct coupling between the Higgs and the inflaton the problem must be cured in a different way. A second possibility is that dissipative effects could generate a non-zero temperature during inflation with would result in corrections to the Higgs mass. That is,
\begin{equation}
V_{\text{eff}} \rightarrow V_{\text{eff}}  +\frac{1}{2} c_h T^2 h^2,
\label{tpot}
\end{equation}
where $c_h \simeq$0.308 in the standard model. Such a temperature might be generated in the context of warm inflation, where the inflaton equation of motion is modified to
\begin{equation}
\ddot{\phi}+(3H+\Upsilon)\dot{\phi}+\frac{dV_\phi}{d\phi}=0,
\end{equation}
where $\Upsilon$ is a model dependent dissipation term that sources a thermal bath. Warm inflation is a well studied subject that tries to use this thermal viscosity to slow the roll of the inflaton and drive inflation \cite{Berera9509}. However for the purposes of the current work we are not interested in the effect of the thermal bath upon the inflaton, but rather on its effects on the Higgs potential. We therefore do not require $\Upsilon$ to be anywhere near as large as $3H$.  Since the functional dependence of $\Upsilon$ on $\phi$ and $T$ is highly model dependent (see e.g. \cite{Berera9803,Berera9809,Berera0808,BasteroGil0902,Yokoyama0909,BasteroGil1008,Mishra1106,LopezNacir1109,BasteroGil1207,Cerezo1210,Bartrum1307}) we do not deal with it directly here. Instead we adopt a completely phenomenological approach and concern ourselves only with the temperature require to stabilise the Higgs field.
\begin{Figure}
\centering
\includegraphics[width=1.1\linewidth]{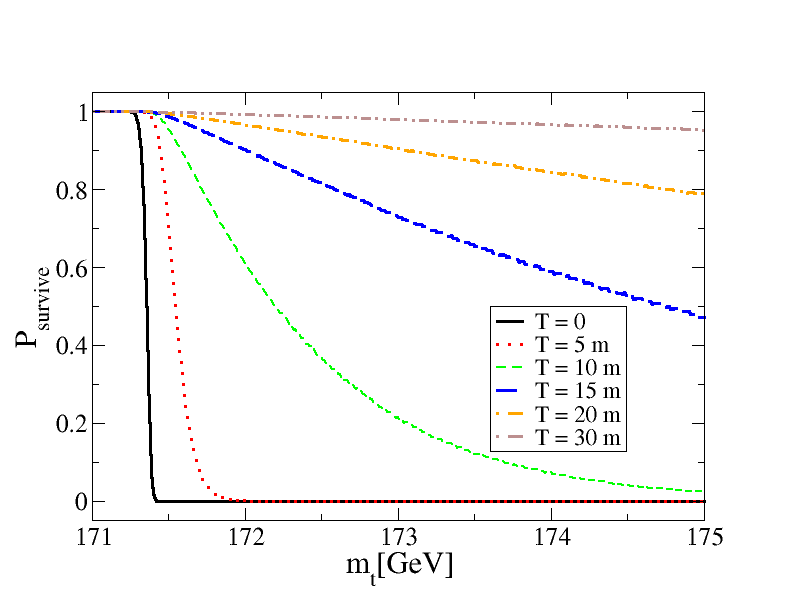}
\captionof{figure}{\it The probability of the Higgs field not ending up above the instability scale $\Lambda$ in any of the Hubble volumes in our past horizon during inflation as a function of top mass $m_t$.  We plot the results for different values of the temperature during inflation $T$.}\label{Temp}
\end{Figure}
Figure \ref{Temp} shows the results of inflation occurring in the presence of a thermal bath of temperature $T$ which prevents the Higgs field from destabilising if it starts with zero expectation value.  Note that we do not require the potential to be completely stable, merely that thermal effects both increase the value of $\Lambda$ via equation (\ref{tpot}) and decrease the variance of the Higgs field $\langle h \rangle$ by changing the effective mass in equation (\ref{eqn: stochastic}).  A temperature of a few tens times the inflaton mass $m$ would therefore prevent the Higgs field from destabilising.

To conclude, in this short letter we have discussed the effect of inflation upon the Higgs field in the light of the exciting possible detection of tensor modes by the BICEP-2 experiment.  If we interpret this as a signal for Inflation with an energy density comparable to the GUT scale then we have shown that even with the most conservative initial condition for the Higgs field, i.e. that it starts at the origin with zero expectation value, fluctuations during inflation will push the field stochastically away from the origin to values that will de-stabilise the electroweak vacuum.  We have shown that this can be avoided by introducing a coupling $\lambda_{\phi h}$ between the Higgs and the inflaton field of the order of a few times $10^{-11}$ or larger. 
 
We have also looked at what temperature would be required during inflation to prevent the Higgs from destabilising and seen that a temperature a few tens times the mass of the inflaton would be sufficient.  It is a bit disappointing that the Gibbons Hawking temperature itself $T_{GH}=H/{2\pi}$ cannot act as the stabiliser of the Higgs potential but it seems to be too small by a couple of orders of magnitude.    

Having knowledge of the energy scale of inflation has radical implications for the thermal history of the Universe, which in turn has a huge bearing upon particle physics.  State of the art calculations have shown that for a Higgs mass of $m_h=125.5$ GeV the time scale for tunneling into the true vacuum is larger than the age of the Universe for top masses up to around $m_t\sim 178$ GeV  \cite{degrassi1205}.  In this work we argue that if the scale of inflation is that suggested by the BICEP-2 experiment then the instability scale for $m_h=125.5$ GeV demands that top quarks must have a mass less than $m_t\sim 172$ GeV (see figure \ref{without}) requiring new physics to step in and protect our vacuum.  This basic work is an example of the kind of physics which we will be able to do in the future if this result is shown to be robust and compatible with the theory of inflation.  Hopefully it is also a small illustration of the monumental implications this discovery, if shown to be correct and consistent with inflation, will have upon our field of research.

\section*{Acknowledgements}
We are extremely grateful for conversations with Eugene Lim and for critical technical assistance from Philipp Grothaus.  MF is grateful for funding provided by the UK Science and Technology Facilities Council. RH is supported by the KCL NMS graduate school. 
\bibliographystyle{h-physrev}
\bibliography{bib_reheat}

\end{document}